
\input harvmac
\input epsf
\noblackbox
\Title{UCLA/95/TEP/10}{The Energy-Momentum Tensor in Field
Theory I $^\star$
\footnote{}{$^\star$ This work was supported in part by the U.S. Department of
Energy, under Contract
DE-AT03-88ER 40384 Mod A006 Task C.}}

\centerline{
Hidenori SONODA$^\dagger$\footnote{}{$^\dagger$
sonoda@physics.ucla.edu}}
\bigskip\centerline{\it Department of Physics and Astronomy,
UCLA, Los Angeles, CA 90024-1547, USA}

\vskip 1in
This is the first of three papers
on the short-distance properties
of the energy-momentum tensor
in field theory.  We study the energy-momentum tensor
for renormalized field theory
in curved space.  We postulate an exact
Ward identity of the energy-momentum tensor.
By studying the consistency of
the Ward identity with the renormalization
group and diffeomorphisms, we determine
the short-distance singularities in the
product of the energy-momentum tensor
and an arbitrary composite field
in terms of a connection for the space
of composite fields over theory space.  We discuss
examples from the four-dimensional $\phi^4$
theory.  In the forthcoming two papers we plan to
discuss the torsion and curvature of the connection.

\Date{April 1995}

\def\H{\Theta}
\def\p{\partial}
\def\L{{\cal L}}
\def\mn{{\mu\nu}}
\def\dt{{d \over dt}~}
\def\e{{\rm e}}
\def\ep{\epsilon}
\def\vev#1{\left\langle #1 \right\rangle_{h,g}}
\def\vvev#1{\left\langle #1 \right\rangle}
\def\K{{\cal K}}
\def\C{{\cal C}}
\def\dO{d^{D-1} \Omega}
\def\O{{\cal O}}
\def\Ct{\tilde{\cal C}}
\def\gone{g_{\bf 1}}
\def\L{{\cal L}}


\newsec{Introduction}

The study of the energy-momentum tensor in field theory has
a long history.  The short-coming of the na{\"\i}ve analysis
of the classical lagrangian in constructing the
energy-momentum tensor, like
the canonical energy-momentum tensor and
Belinfante tensor, was noticed long ago.
(See ref.~\ref\rjackiw{R.~Jackiw, {\it in}
Lectures on Current Algebra and Its Applications,
(Princeton U.P., 1972)} and references therein
for a convenient summary of the earlier works.)
The purpose of the present paper
and its sequels is to give a fresh look at this old
subject of the energy-momentum tensor and its
short-distance singularities.

Our motivation for the present study comes
partially from string field theory.  One of the most
important issues in perturbative string field theory
is to formulate the theory using an arbitrary
two dimensional non-linear sigma model as a background.
(Here, the non-linear sigma field takes values
on an arbitrary space-time manifold, and the model
has an infinite number of parameters including the space-time
metric and dilaton field.)
This is necessary for any discussion of background independence
to make sense at all.  Our goal is, then,
to express the dependence of the theory on the
parameters and the world-sheet metric in a way
suitable for string field theory.

The dependence of renormalized field theories
on their parameters has been studied recently
in the framework of the variational
formula \ref\rsonoda{H.~Sonoda,
{\it Nucl. Phys.} {\bf B383}(1992)173;
``Connection on the theory space,''
talk given at Strings 93 conference, Berkeley, May 1993,
(hep-th/9306119)}.  The notion of a field $\O_i$
conjugate to a parameter $g^i$ has been introduced,
and the spatial integral over the conjugate field has been
given a precise meaning.  In doing so, we have introduced
a connection $c_i$ for the linear space of composite fields.
The short-distance singularities in the product of a conjugate
field and an arbitrary composite field are expressed in terms
of the scale dimension of the
composite field and the connection $c_i$.  This formulation
of field theory in coordinate space,
especially its application to conformal field
theory \ref\rrsz{K.~Ranganathan,
H.~Sonoda, and B.~Zwiebach, {\it Nucl. Phys.}
{\bf B414}(1994)405},
has turned out to be
the necessary tool in proving the background
independence of string
field theory based upon a conformal field theory with
continuous parameters \ref\rsz{A.~Sen and B.~Zwiebach,
{\it Nucl. Phys.} {\bf B414}(1994)649; {\bf B423}(1994)580}.

We will study the energy-momentum tensor
using the techniques similar to the above variational formula.
It is possible to consider the energy-momentum
tensor strictly in the realm of field theory in flat space.
But, as was pointed out for the first time in
ref.~\ref\rccj{C.~G.~Callan, S.~Coleman,
and R.~Jackiw, {\it Ann. Phys.} {\bf 59}(1970)42},
the energy-momentum tensor is best
defined in curved space, or, equivalently,
it can be defined unambiguously if we couple the theory
to external gravity.  Hence, we study general renormalized
field theories in curved space, i.e., a Riemann manifold
with metric $h_\mn$.  We will define
the energy-momentum tensor as
the field that is conjugate to the metric $h_\mn$.

We are interested in the general properties of the
energy-momentum tensor.  We would like to isolate
those features of the energy-momentum tensor
that specify it uniquely.  We would also like
to study the short-distance singularities of the
product of the energy-momentum tensor
and an arbitrary composite field.
We wish, in the course of this work,
to take up the old subject which was initiated in
refs.~\rccj,
\ref\rcj{S.~Coleman and R.~Jackiw, {\it Ann. Phys.} {\bf 67}(1971)552}, and
\ref\rwilson{K.~Wilson, {\it Phys. Rev.} {\bf D2}(1970)1478}.
The reader will recognize a fuller use
of the renormalization group (RG) in our study.

In sect.~2 we will postulate
an exact Ward identity for the energy-momentum tensor.
In writing the Ward identity, the short-distance singularities
are subtracted carefully to obtain a finite result.
We will show that the finite counterterms in the Ward identity
can be interpreted as the matrix elements of
a connection for the linear
space of composite fields over theory space.
In sects.~3, 4, and 5 we check the consistency of the Ward
identity postulated in sect.~2.  In sect.~3 we will demand that the Ward
identity
is consistent with the RG.
Consequently
we will find that the energy-momentum tensor has no anomalous
dimension and that the short-distance singularities in the product
of the energy-momentum tensor and an arbitrary composite field
are determined by the connection introduced in sect.~2.
In sect.~4 we will see
that the consistency of the Ward identity with
the variational formula that determines how
the theory depends on the parameters gives a derivation
of the famous trace anomaly.   We will obtain an expression
of the trace of the energy-momentum tensor
as a sum over the fields conjugate to the parameters.
In sect.~5 we will require
consistency with diffeomorphisms.  We will write
down the euclidean analogue of the commutator between the
energy-momentum tensor and an arbitrary composite field
in terms of the connection of sect.~2.
In sect.~6 we will study further the short-distance singularities
of the energy-momentum tensor, found in sects.~3 and 5.
In sect.~7 we will discuss the characteristic properties of the energy-momentum
tensor which specify it unambiguously.
In sect.~8 we will discuss $\phi^4$
theory in four dimensional flat space to elucidate the
general discussions of the preceding sections.
We give concluding remarks in sect.~9.

\newsec{The exact Ward identity of the energy-momentum tensor}

We consider a renormalized field theory with renormalized
parameters $g^i (i=1,...,N)$ on a $D$-dimensional manifold
with a positive definite metric $h_\mn$.  We regard the parameters
$g^i$ and the metric $h_\mn$ as local
coordinates of the theory space.

We introduce the renormalization group (RG) transformation
on the theory space as
\eqna\eRG
$$\eqalignno{\dt g^i &= \beta^i (g) &\eRG a\cr
\dt h_\mn &= - 2 h_\mn ~.&\eRG b\cr}
$$
Note that eq.~\eRG{b}\ implies that the physical
distance between two points $r^\mu$ and $r^\mu +
\delta r^\mu$ goes as $\e^{- t}$ under the RG
transformation.

Let $\{\Phi_a\}$ be a basis of composite fields.  We only
take covariant local fields for simplicity.  We can
define a new basis by
\eqn\ebasis{
\Phi'_a = \left( N (h,g) \right)_a^{~b} \Phi_b~,}
where $N(h,g)$ is an invertible matrix which can
depend on both the parameters $g^i$ and the metric
$h_\mn$, curvature, and its covariant derivatives.
The composite fields satisfy the RG equations
\eqn\eRGfield{ \dt \Phi_a = \left( \gamma (h,g) \right)_a^{~b}
\Phi_b ~,}
where $\gamma$ is the matrix of the full scale dimension,
in the sense that the correlation function of $n$ arbitrary
composite fields satisfies the following RG equation:
\eqn\eRGeq{\eqalign{
&\dt \vev{\Phi_{a_1} (P_1) ... \Phi_{a_n} (P_n)}\cr
& \quad \equiv \lim_{\Delta t \to 0} {1 \over \Delta t}
\left[ \vvev{\Phi_{a_1} (P_1) ... \Phi_{a_n} (P_n)}_{
(1 - 2 \Delta t) h, g + \Delta t \beta} -
\vev{\Phi_{a_1} (P_1) ... \Phi_{a_n} (P_n)} \right]\cr
& \quad = \sum_{k=1}^n \left[ \gamma (h(P_k), g)
\right]_{a_k}^{~b}
\vev{\Phi_{a_1} (P_1) ... \Phi_b (P_k) ... \Phi_{a_n} (P_n)}~.\cr}}
We note that the convention of the RG equations
adopted here differs somewhat from the convention
adopted on flat space for which the metric is
\eqn\eflat{h_\mn (r) = \delta_\mn~.}
To keep the flat metric invariant under the RG,
we must compensate eq.~\eRG{b}\ by a coordinate
transformation
\eqn\ecoord{ r^\mu \to (1 - \Delta t) r^\mu~,}
where $\Delta t$ is an infinitesimal change of the
scale parameter $t$.  If the field $\Phi_a$ is a
tensor of rank (m,n), it transforms as
\eqn\etransPhi{(\Phi_a)_{\mu_1 ... \mu_m}^{\nu_1 ... \nu_n}
\to ( 1 + (m-n) \Delta t)
(\Phi_a)_{\mu_1 ... \mu_m}^{\nu_1 ... \nu_n}}
under \ecoord.  Hence, in the flat space the RG transformation
is given by eqs.~\eRG{a}\ for the parameters and
\eqn\eRGflat{\dt \Phi_a = (m-n) \Phi_a + \left(
\gamma (\delta,g) \right)_a^{~b} \Phi_b~.}
for the composite fields.

We define the energy-momentum tensor $\H^\mn$ as
the composite field that generates infinitesimal changes
of the metric tensor $h_\mn$.  More specifically we assume
the existence of
$\H^\mn$ that satisfies the following
exact Ward identity:
\eqnn\eWard
$$
\eqalignno{&\vev{\Phi_{a_1} (P_1) ... \Phi_{a_n} (P_n)}
- \vvev{\Phi_{a_1} (P_1) ... \Phi_{a_n} (P_n)}_{h + \delta h, g}\cr
& = \lim_{\epsilon \to 0} \Bigg[
\int_{\rho (r,P_k) \ge \epsilon} d^D r \sqrt{h} ~{1 \over 2} \delta
h_\mn (r) \cr
& \qquad \qquad \qquad
\times \vev{ \left( \H^\mn (r) - \vev{\H^\mn (r)} \right)
\Phi_{a_1} (P_1) ... \Phi_{a_n} (P_n)} \cr
& \quad + \sum_{k=1}^n
\sum_{m=0}^\infty {1 \over m!} \nabla_{\mu_1} ... \nabla_{\mu_m}
{1 \over 2} \delta h_\mn (P_k) &\eWard\cr
& \qquad\quad \times \left\{
\left( \K^{\mn, \mu_1 ... \mu_m} (h(P_k),g) \right)_{a_k}^{~b}
- \int_\epsilon^1 d\rho
\left( \C^{\mn, \mu_1 ... \mu_m} (\rho; h(P_k),g) \right)_{a_k}^{~b} \right\}
\cr
& \qquad \qquad \qquad \times
\vev{\Phi_{a_1} (P_1) ...  \Phi_b (P_k) ... \Phi_{a_n} (P_n)}
\quad\Bigg]~,
\cr }
$$
where $\delta h_\mn$ is an infinitesimal symmetric tensor.
The Ward identity specifies only the symmetric part
of $\H^\mn$, and we can define $\H^\mn$ to be symmetric.
The symbol $\rho(r, P)$ denotes the geodesic distance
between the points $r$ and $P$.  We must exclude
infinitesimal balls $\rho(r,P_k) \le \ep$ from the
domain of integration since the product
$\H^\mn (r) \Phi_{a_k} (P_k)$ contains short-distance
singularities.

Let us elaborate on the coefficients $\C^{\mn, \mu_1 ... \mu_m}$
in the exact Ward identity \eWard.
In a neighborhood of a point $P$ we can
decompose the volume element $\sqrt{h(r)} d^D r$ to
the product of the volume elements for the radial and angular
parts:
\eqn\evolume{\sqrt{h(r)} d^D r = d\rho ~\dO_\rho (r,P) ~.}
Then, given an arbitrary symmetric tensor $t_\mn (r)$,
we can expand the product $\H^\mn (r) \Phi_a (P)$ as
\eqn\eOPE{\eqalign{&\int \dO_\rho (r,P) ~t_\mn (r)
\H^\mn (r) \Phi_a (P)\cr
& = \sum_{m=0}^\infty {1 \over m!}
\nabla_{\mu_1} ... \nabla_{\mu_m}
t_\mn (P) \cdot \left( \C^{\mn, \mu_1 ... \mu_m} (\rho; h(P), g)
\right)_a^{~b}
\Phi_b (P) + {\rm o} \left( {1 \over \rho (r,P)} \right)~,\cr}}
where we only keep the part which cannot be integrated
over $\rho$ to the origin.  Because
of this, the sum over the integer $m$ is a finite sum.
The coefficients $\left( \C^{\mn, \mu_1 ... \mu_m} \right)_a^{~b}$
is a tensor at the point $P$ which can depend on the
geodesic distance $\rho (r,P)$, metric
$h_\mn (P)$, curvature $R_{\mu \nu \alpha}^{~~~~\beta} (P)$,
its covariant derivatives at $P$, and parameters $g^i$.
The coefficient $\C^{\mn, \mu_1 ... \mu_m}$
is symmetric with respect to $\mu_1, ... , \mu_m$.

Coming back to the exact Ward identity \eWard, we can
take the limit $\ep \to 0$ thanks to the subtraction of the
short-distance singularities \eOPE.  To compensate the
arbitrariness of the subtraction, we need
to introduce finite counterterms $\K^{\mn, \mu_1 ... \mu_m}$.

It is easy to see that the finite counterterms behave as the
matrix elements of a connection for the linear space of composite
fields over theory space.  Under the change of basis
\ebasis, we find that the operator product expansion (OPE)
coefficients transform covariantly as
\eqn\eCtransf{(\delta h \cdot \C) (\rho; h(P),g)
\to N(h(P),g) \cdot (\delta h \cdot \C) (\rho; h(P),g)
\cdot  N^{-1} (h(P),g) ~,}
but the finite counterterms transform as
\eqn\eKtransf{\eqalign{& (\delta h \cdot \K) (h(P),g) \to
N(h(P),g) \cdot (\delta h \cdot \K) (h(P),g) \cdot N^{-1} (h(P),g)\cr
&\quad + \left( N(h(P),g) - N(h+\delta h(P),g) \right)
N^{-1} (h(P),g) ~,\cr}}
where we have introduced the short-hand notation
\eqn\edeltahCK{
\eqalign{
(\delta h \cdot \C) (h(P),g) &\equiv
\sum_{m=0}^\infty {1 \over m!}
\nabla_{\mu_1} ... \nabla_{\mu_m} {1 \over 2}~\delta h_\mn (P) \cdot
\C^{\mn, \mu_1 ... \mu_m} (h(P),g)~,\cr
(\delta h \cdot \K) (h(P),g)
&\equiv \sum_{m=0}^\infty {1 \over m!}
\nabla_{\mu_1} ... \nabla_{\mu_m} {1 \over 2}~\delta h_\mn (P) \cdot
\K^{\mn, \mu_1 ... \mu_m} (h(P),g) ~.\cr}}
{}From the transformation property \eKtransf, we see that
$\K^{\mn, \mu_1 ... \mu_m}$ is a connection for the linear
space of composite fields over theory space.

Here we should recall that the
theory space has the metric $h_\mn$ and parameters
$g^i$ as local coordinates.
Therefore, strictly speaking,
$\K^{\mn, \mu_1 ... \mu_m}$ gives the
connection in the direction of the metric deformation
on the theory space.
The elements of the connection in the direction of
the parameters $g^i$ have been introduced as
$c_i$
in ref.~\rsonoda.  Though only field theory
in flat space is discussed in this reference,
the generalization to curved space is straightforward.

For the reader's convenience, let us summarize
the results of ref.~\rsonoda\ relevant to this paper.
The connection $c_i$ has been introduced
as finite counterterms in the variational formula that
expresses how the correlation functions change under
infinitesimal changes of the parameters $g^i$:
\eqnn\evar
$$
\eqalignno{
&- {\partial \over \partial g^i} \vev{\Phi_{a_1} (P_1) ...
\Phi_{a_n} (P_n)} \cr
& = \int_{\rho (r,P_k) \ge \ep} d^D r \sqrt{h}~\vev{
\left( \O_i (r) - \vev{\O_i (r)} \right)
\Phi_{a_1} (P_1) ... \Phi_{a_n} (P_n)} \cr
&\qquad + \sum_{k=1}^n \left[
(c_i)_{a_k}^{~b} (h(P_k),g) - \int_\ep^1 d\rho
\left( \C_i \right)_{a_k}^{~b} (\rho;h(P_k),g) \right]
&\evar\cr
&\qquad \qquad \qquad
\times \vev{\Phi_{a_1} (P_1) ... \Phi_b (P_k) ... \Phi_{a_n} (P_n)}~,\cr}
$$
where $\O_i$ is the composite field conjugate
to the parameter $g^i$, and the OPE
coefficients $\C_i$ are defined by
\eqn\eOPEi{
\int \dO_\rho~\O_i (r) \Phi_a(P)
= \left( \C_i \right)_a^{~b} (\rho; h(P),g) \Phi_b (P) +
{\rm o} \left( {1 \over \rho} \right)~.}
Note that the connection $c_i (h(P),g)$ in general depends not only
on the parameters but also on the metric, curvature, and
its derivatives at point $P$.  Under the change of basis \ebasis,
the connection $c_i$ transforms as
\eqn\ectransf{c_i (h(P),g) \to
N(h(P),g) \left( c_i (h(P),g) + {\partial \over \partial g^i}
\right) N^{-1} (h(P),g)~.}
In general the conjugate field $\O_i$ is ambiguous up to
a total derivative $\nabla_\mu J_i^\mu$, but we assume that
we can remove the ambiguity by demanding the absence of
mixing with total derivatives under the RG.  Then,
the conjugate field $\O_i$ satisfies the RG equation
\eqn\eRGconj{\dt \O_i = D \O_i - {\partial \beta^j \over \partial
g^i} ~\O_j~.}
Finally we note that the consistency of
the variational formula \evar\ with the RG gives
the OPE coefficients $\C_i$ as
\eqn\eCi{\C_i (1;h,g) = \p_i \gamma
+ \dt c_i - [\gamma, c_i] + \p_i \beta^j \cdot c_j ~.}

The purpose of the following three sections is to
check consistency of the exact Ward identity \eWard.

\newsec{Consistency with the RG}

We demand that the exact Ward identity
\eWard\ be consistent with the RG equations \eRG{}.
In ref.~\rsonoda, it is shown that the consistency of
the variational formula \evar\ with the RG gives rise to
the expression \eCi\ of the OPE coefficients $\C_i$.
We proceed analogously here.

We can compute
\eqn\eDelta{\Delta \equiv
\vvev{\Phi_a (P)}_{\e^{- 2 \Delta t} (h + \delta h),
g + \Delta t \beta} - \vev{\Phi_a (P)}}
to first order in $\Delta t$ and $\delta h$ in two different
ways.  The results must agree.
The two methods are shown schematically in Fig. 1.
\smallskip
\centerline{\epsfxsize=0.5\hsize \epsfbox{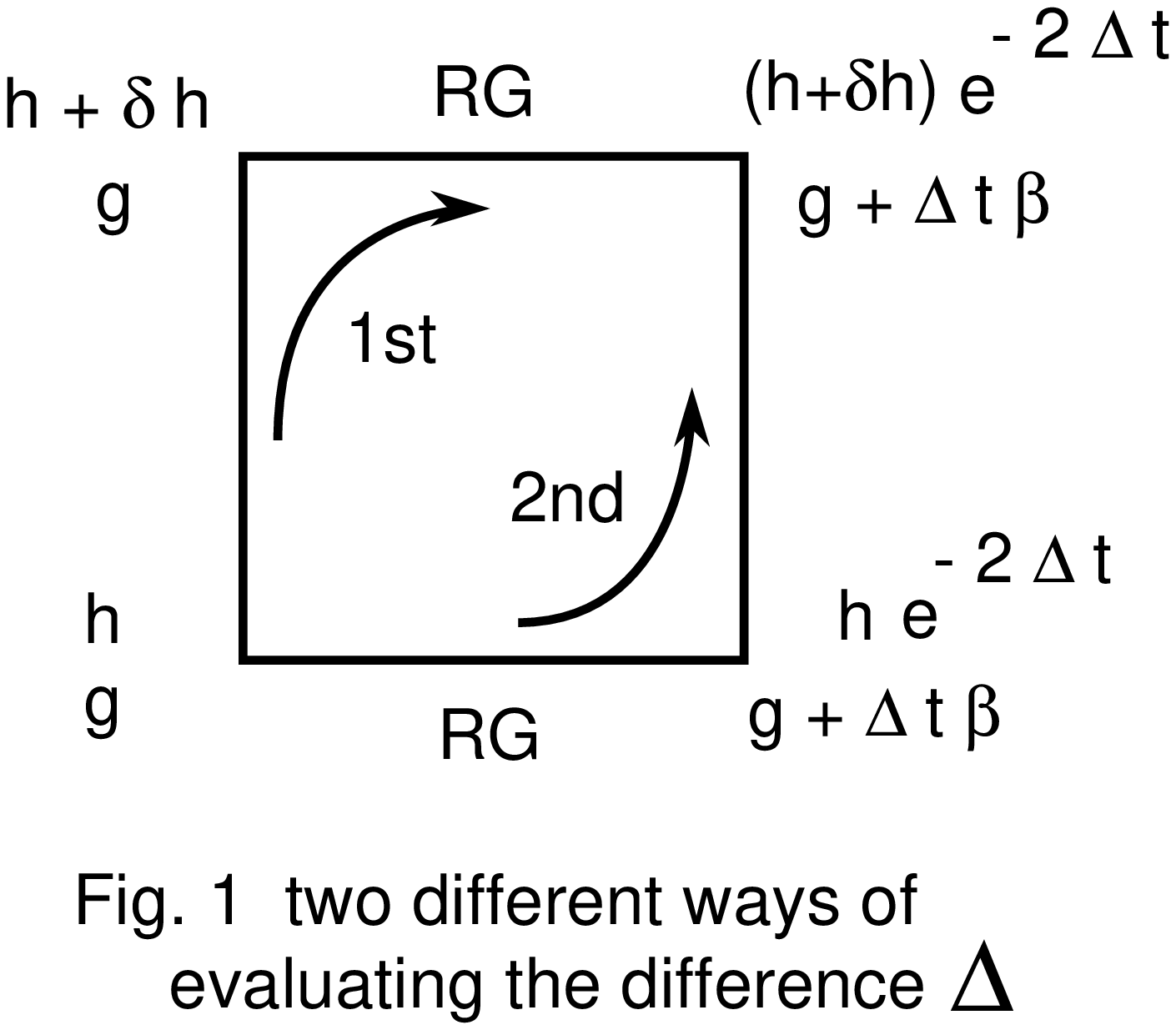}}

In the first method we calculate
\eqn\emethodia{\eqalign{\Delta
&= \left( \vvev{\Phi_a (P)}_{(h+\delta h)\e^{-2 \Delta t},
g + \Delta t \beta} - \vvev{\Phi_a (P)}_{h + \delta h, g} \right) \cr
&\quad + \left( \vvev{\Phi_a (P)}_{h + \delta h, g}
- \vev{\Phi_a (P)} \right)~. \cr}}
We apply the RG eq.~\eRGeq\ to the first bracket
and obtain
\eqn\emethodib{\eqalign{
\Delta &= \Delta t~\gamma_a^{~b} (h+\delta h, g)
\vvev{\Phi_b (P)}_{h+\delta h, g} \cr
&\quad + \left( \vvev{\Phi_a (P)}_{h + \delta h, g}
- \vev{\Phi_a (P)} \right)~.\cr}}
By applying the exact Ward identity \eWard, we find
\eqnn\emethodic
$$
\eqalignno{
\Delta &= \Delta t~\gamma_a^{~b} (h,g) \vev{\Phi_b (P)}
+ \left( \vvev{\Phi_a (P)}_{h + \delta h, g}
- \vev{\Phi_a (P)} \right) \cr
&\quad + \Delta t~\delta_h \gamma_a^{~b} (h,g) \vev{\Phi_b (P)}
&\emethodic\cr
& \quad + \Delta t ~\gamma_a^{~b} (h,g) \Bigg[
 - \int_{\rho(r,P) \ge \ep} d^D r \sqrt{h} ~{1 \over 2}
\delta h_\mn (r) \vev{ \H^\mn (r) \Phi_b (P)}^c \cr
&\quad \quad + \left(
\int_\ep^1 d\rho~\delta h \cdot \C (\rho;h,g)
- \delta h \cdot \K (h,g) \right)_b^{~c}
\vev{\Phi_c (P)} \Bigg]~,
\cr}
$$
where $^c$ denotes the connected part, and
\eqn\edgamma{\delta_h \gamma (h,g) \equiv
\gamma (h+\delta h, g) - \gamma (h,g)~.}

In the second method we calculate
\eqn\emethodiia{\eqalign{\Delta &= \left(
\vvev{\Phi_a (P)}_{(h+\delta h)\e^{- 2 \Delta t}, g + \Delta t \beta}
- \vvev{\Phi_a (P)}_{h\e^{- 2 \Delta t}, g + \Delta t \beta} \right)\cr
&\quad + \left( \vvev{\Phi_a (P)}_{h\e^{- 2 \Delta t}, g + \Delta t \beta} -
\vev{\Phi_a (P)} \right)~.\cr}}
Using the Ward identity \eWard\ and the RG eq.~\eRGeq,
we obtain
\eqnn\emethodiib
$$
\eqalignno{\Delta & = - \int_{\rho(r,P; h\e^{- 2 \Delta t})
\ge \ep \e^{- \Delta t}} d^D r \sqrt{h} ~\e^{- D \Delta t}\cr
&\qquad \qquad \qquad \times {1 \over 2}
\delta h_\mn (r) \e^{- 2 \Delta t} \vvev{ \H^\mn (r) \Phi_a (P)}_{
h \e^{- 2 \Delta t}, g + \Delta t \beta}^c \cr
&\quad + \Bigg[ \int_{\ep \e^{- \Delta t}}^1
d \rho~\e^{- 2 \Delta t} \delta h \cdot
\C (\rho; h\e^{- 2 \Delta t}, g + \Delta t \beta)
&\emethodiib\cr
&\quad\quad - \e^{- 2 \Delta t} \delta h \cdot
\K ( h\e^{- 2 \Delta t}, g + \Delta t \beta) \Bigg]_a^{~b}
\vvev{\Phi_b (P)}_{h \e^{- 2 \Delta t}, g + \Delta t \beta} \cr
&\quad + \Delta t~\gamma_a^{~b} (h,g) \vev{\Phi_b (P)}~.\cr}
$$
Applying the RG eq.~\eRGeq, we find
\eqnn\emethodiic
$$
\eqalignno{&\Delta =
\Delta t~\gamma_a^{~b} (h,g) \vev{\Phi_b (P)}
+ \left( \vvev{\Phi_a (P)}_{h + \delta h, g}
- \vev{\Phi_a (P)} \right) \cr
& + \Delta t~\Bigg[
- \gamma_a^{~b} (h,g) \int_{\rho (r,P)\ge \ep}
d^D r \sqrt{h}~{1 \over 2}
\delta h_\mn (r)~\vev{\H^\mn (r) \Phi_b (P)}^c \cr
&\quad - \int_{\rho (r,P) \ge \ep}
d^D r \sqrt{h}~{1 \over 2} \delta h_\mn (r) \cr
&\qquad \qquad \qquad \times \vev{ \left( \dt \H^\mn (r) - (D+2) \H^\mn (r)
\right)
\Phi_a (P)}^c &\emethodiic\cr
&\quad +
\left( \delta h \cdot \C (1;h,g) \right)_a^{~b} \vev{\Phi_b (P)}\cr
&\quad + \int_\ep^1
d\rho~ \left\{ \delta h \cdot \left(
\dt \C (\rho; h,g)  + \C (\rho;h,g) \gamma (h,g) \right)
\right\}_a^{~b} \vev{\Phi_b (P)}\cr
&\quad - \left\{ \delta h \cdot
\left( \dt \K (h,g) + \K (h,g) \gamma (h,g) \right) \right\}_a^{~b}
\vev{\Phi_b (P)}~\Bigg]~,\cr}
$$
where
\eqn\edtCdtK{\eqalign{
\dt \C (\rho;h,g) &\equiv {1 \over \Delta t}
\left[ \e^{- 3 \Delta t} \C (\rho \e^{- \Delta t}; h \e^{- 2 \Delta t},
g + \Delta t \beta) - \C (\rho; h,g) \right] \cr
\dt \K (h,g) &\equiv {1 \over \Delta t}
\left[ \e^{- 2 \Delta t} \K (\rho \e^{- \Delta t}; h \e^{- 2 \Delta t},
g + \Delta t \beta) - \K (\rho; h,g) \right] ~.\cr}}

Now, eq.~\emethodiic\ must agree with eq.~\emethodic.
This gives
\eqnn\erelation
$$
\eqalignno{
&\left[ \delta_h \gamma +
\delta h \cdot \left\{ \gamma (h,g)
\left( \int_\ep^1 d\rho ~\C (\rho;h,g) - \K (h,g) \right) \right\}
\right]_a^{~b}
\vev{\Phi_b (P)} \cr
& = - \int_{\rho (r,P) \ge \ep}
d^D r \sqrt{h}~{1 \over 2} \delta h_\mn (r) \cr
&\qquad \qquad \quad \times \vev{ \left( \dt \H^\mn (r) -
(D+2) \H^\mn (r) \right)
\Phi_a (P)}^c &\erelation\cr
&~+ \Bigg[  \delta h \cdot \Bigg\{
\C (1;h,g) + \int_\ep^1 d\rho~\left(
\dt \C (\rho; h,g)  + \C (\rho;h,g) \gamma (h,g) \right) \cr
&\qquad \qquad\qquad - \dt \K (h,g) -
\K (h,g) \gamma (h,g) \Bigg\} \Bigg]_a^{~b}
\vev{\Phi_b (P)}~.\cr}
$$
Since the left-hand side is local, the integrand of
the integral on the
right-hand side must be a total derivative.
For this to be valid for an arbitrary $\delta h_\mn$,
we must find
\eqn\eEMRG{\dt \H^\mn = (D+2) \H^\mn~.}
This implies that the energy-momentum tensor has no
anomaly under the renormalization group.  The RG equations
\eRGeq\ and \eEMRG\ imply that
\eqn\eCRG{\dt \C (\rho; h,g) = \left[ \gamma (h,g) ,
\C (\rho; h,g) \right]~,}
where the square bracket denotes a commutator.
Substituting eqs.~\eEMRG\ and \eCRG\ into
eq.~\erelation, we obtain
\eqn\esing{ \delta h \cdot
\C (1;h,g)  = \delta_h \gamma (h,g)
+ \delta h \cdot
\left( \dt \K (h,g) - [\gamma (h,g) , \K (h,g) ] \right)~.}
This gives the OPE coefficients $\C$ in terms of the
scale dimensions $\gamma$ and connection
$\K$ in the same way that eq.~\eCi\ gives the OPE
coefficients $\C_i$ in terms of the scale dimensions
$\gamma$ and connection $c_i$.  The relation \esing\ is a main
result of this paper.

\newsec{Consistency with the variational formula}

In this section we demand consistency between
the exact Ward identity \eWard\ and the variational
formula \evar.  We recall the RG equation \eRGeq\
for $n=1$:
\eqn\eRGeqii{\vvev{\Phi_a (P)}_{(1-2 \Delta t) h,g+\Delta t \beta}
- \vev{\Phi_a (P)} = \Delta t ~\gamma_a^{~b} (h(P), g) \vev{\Phi_b
(P)}~.}
We can evaluate the left-hand side by using the
exact Ward identity \eWard\ and the variational
formula \evar.  Keeping only terms of first order
in $\Delta t$, we obtain
\eqn\eWardvari{\eqalign{
&\vvev{\Phi_a (P)}_{(1-2 \Delta t) h, g+\Delta t \beta}
- \vev{\Phi_a (P)} \cr
& = \left( \vvev{\Phi_a (P)}_{(1-2 \Delta t) h,g}
- \vev{\Phi_a (P)} \right) +
\left( \vvev{\Phi_a (P)}_{h,g+\Delta t \beta}
- \vev{\Phi_a (P)} \right) ~.\cr}}
We use \eWard\ for the first bracket and \evar\ for the second
to obtain
\eqnn\eWardvarii
$$
\eqalignno{&\vvev{\Phi_a (P)}_{(1-2 \Delta t) h,
g+\Delta t \beta}
- \vev{\Phi_a (P)} \cr
& = \Delta t \Bigg[ \int_{\rho(r,P) \ge \ep}
d^D r \sqrt{h} ~ \vev{ \left( \H (r)
- \beta^i \O_i (r) \right) \Phi_a (P)}^c &\eWardvarii\cr
& \quad + \int_\ep^1 d\rho~ \left(
- h_\mn (P) \C^\mn + \beta^i \C_i \right)_a^{~b}
(\rho; h (P),g) \vev{\Phi_b (P)} \cr
& \quad + \left( h_\mn (P) \K^\mn - \beta^i c_i \right)_a^{~b}
(h(P),g) \vev{\Phi_b (P)} \Bigg] ~,\cr}
$$
where we denote the trace of the energy-momentum tensor by
\eqn\eH{\H \equiv h_\mn \H^\mn~.}
For eqs.~\eRGeqii\ and \eWardvarii\ to agree, the difference
$\H - \beta^i \O_i$ must be a total derivative
\eqn\ediff{\H - \beta^i \O_i = \nabla_\mu J^\mu}
so that the integral in \eWardvarii\ reduces to local terms
at $P$.  From \eEMRG\ and \eRGconj, however, both
the trace $\H$ and $\beta^i \O_i$ have canonical
dimension $D$ under the RG, and we must find
\eqn\eRGdJ{\dt \nabla_\mu J^\mu = D \nabla_\mu J^\mu~.}
This implies that the current $J^\mu$ has no anomalous
dimension except that it may mix with a conserved current
$j^\mu$:
\eqn\eRGJ{\dt J^\mu = D J^\mu + j^\mu~.}
We assume that such a non-conserved current
$J^\mu$ does not exist.  Therefore, we conclude that
\eqn\etrace{\H = \beta^i \O_i~.}
This is the well-known trace anomaly.
It means that the sum of the conjugate fields
$\beta^i \O_i$ generates global scale transformations.

Finally, the consistency between
eqs.~\eRGeqii\ and \eWardvarii\ gives a constraint
on the trace of the connection:
\eqn\etraceK{\eqalign{
&h_\mn (P) \K^\mn (h(P),g) = 2 h (P) \cdot \K (h(P),g) \cr
&\quad = \Psi (h(P),g)
\equiv \gamma (h(P),g) + \beta^i c_i (h(P),g)~.\cr}}
This turns out to be very useful for
practical applications as we will see in sect.~8.

\newsec{Consistency with diffeomorphisms}

The most interesting consistency check
is that with diffeomorphisms.  We can express
an arbitrary infinitesimal diffeomorphism by
\eqn\ediffeo{h_\mn \to h_\mn + \left(\L_u h\right)_\mn ~,}
where $\L_u$ denotes the Lie derivative along
an infinitesimal vector field $u^\mu$:
\eqn\eLieh{\L_u h_\mn \equiv \nabla_\mu u_\nu + \nabla_\nu
u_\mu~.}
Under the diffeomorphism we must find
\eqn\ediffeoPhi{\vvev{\Phi_a (P)}_{h + \L_u h, g}
= \vev{\Phi_a (P)} + \vev{\L_u \Phi_a (P)}~.}
Recall that if $\Phi$ is a tensor of rank (m,n), its
Lie derivative is defined by
\eqn\eLie{\L_u \Phi_{\mu_1 ... \mu_m}^{\nu_1 ... \nu_n}
\equiv u^\mu \p_\mu \Phi_{\mu_1 ... \mu_m}^{\nu_1 ... \nu_n}
+ \sum_{k=1}^m \p_{\mu_k}
u^\mu~\Phi_{\mu_1 ... \mu ... \mu_m}^{\nu_1 ... \nu_n}
- \sum_{k=1}^n \p_\nu
u^{\nu_k}~\Phi_{\mu_1 ... \mu_m}^{\nu_1 ... \nu ... \nu_n}~.}

We can also calculate the change of the expectation value
under the diffeomorphism
using the Ward identity \eWard.  We find
\eqn\ediffi{\eqalign{
&\vvev{\Phi_a (P)}_{h + \L_u h, g} - \vev{\Phi_a (P)} \cr
&\quad = - \int_{\rho (r,P) \ge \ep} d^D r \sqrt{h}~\nabla_\mu
u_\nu (r) \vev{\H^\mn (r) \Phi_a (P)}^c \cr
&\quad\quad + \left[ \L_u h \cdot
\left( \int_\ep^1 d\rho~\C (\rho;h,g) - \K (h,g) \right) \right]_a^{~b}
\vev{\Phi_b (P)}~.}}
Integration by parts gives
\eqnn\ediffii
$$
\eqalignno{
&\vvev{\Phi_a (P)}_{h + \L_u h, g} - \vev{\Phi_a (P)} \cr
&\quad = \int_{\rho (r,P) \ge \ep} d^D r \sqrt{h}~u_\nu (r) \vev{
\nabla_\mu \H^\mn (r) \Phi_a (P)}^c \cr
&\quad \quad + \int_{\rho (r,P) = \ep} \dO_\ep~N_\mu (r)
u_\nu (r) \vev{\H^\mn (r) \Phi_a (P)}^c &\ediffii\cr
&\quad\quad + \left[ \L_u h \cdot
\left( \int_\ep^1 d\rho~\C (\rho;h,g) - \K (h,g) \right) \right]_a^{~b}
\vev{\Phi_b(P)} ~,\cr}
$$
where $N^\mu (r)$ is the outward normal vector of unit length
at a point $r$ on the sphere $\rho (r,P) = \ep$.
Eq.~\ediffii\ must agree with eq.~\ediffeoPhi.  Hence, the integrand
of the volume integral
on the right-hand side of \ediffii\ must be a total derivative.
Since the vector field $u^\mu$ is arbitrary,
this condition implies that the energy-momentum tensor satisfies
the conservation law:
\eqn\econserv{\nabla_\mu \H^\mn = 0~.}
This is a well known result.

To proceed further we need some preparation
on local riemannian geometry.  We go back to eq.~\eOPE,
the definition of the OPE coefficients $\C$.  The tensor field
$t_\mn (r)$ is regular at $P$, and we can Taylor-expand
it as
\eqn\eTaylor{t_\mn (r)
= \sum_{m=0} {1 \over m!} v^{\mu_1} (r) ... v^{\mu_m} (r) \cdot
\nabla_{\mu_1} ... \nabla_{\mu_m} t_{\alpha\beta} (P) \cdot
V^\alpha_{~\mu} (P,r) V^\beta_{~\nu} (P,r)~,}
where $v^\mu (r)$ is a tangent vector at $P$ such that its
image under the exponential map at $P$ is the point $r$, i.e.,
\eqn\eexp{Exp_v (P) = r~,}
and $V^\mu_{~\alpha} (P,r)$ is the operator that
parallel transports vectors at $r$ to vectors at $P$
along the geodesic between the two points.
The covariant derivatives are symmetrized automatically
in \eTaylor.  Using the Taylor expansion
\eTaylor, we can write
\eqn\eOPEii{
\eqalign{&\int \dO_\rho (r,P) ~t_\mn (r)
\H^\mn (r) \Phi_a (P)
= \sum_{m=0}^\infty {1 \over m!}
\nabla_{\mu_1} ... \nabla_{\mu_m}
t_{\alpha\beta} (P)\cr
&\quad \times \int \dO_\rho (r,P) ~v^{\mu_1} (r) ...
v^{\mu_m} (r) V^\alpha_{~\mu} (P,r)
V^\beta_{~\nu} (P,r) \H^\mn (r) \Phi_a (P)~.\cr}}
Comparing this with the definition \eOPE\ of the
OPE coefficients, we obtain a relation
\eqn\eCintegral{\eqalign{&\int \dO_\rho (r,P) ~v^{\mu_1} (r) ...
v^{\mu_m} (r) V^\alpha_{~\mu} (P,r)
V^\beta_{~\nu} (P,r) \H^\mn (r) \Phi_a (P) \cr
&\quad = \left(\C^{\alpha\beta, \mu_1 ... \mu_m} (\rho;h(P),g)
\right)_a^{~b} \Phi_b (P)
+ {\rm o} \left( {1 \over \rho (r,P)} \right) \cr}}
that we will use later.  The symmetry with respect to
$\mu_1, ... , \mu_m$ is manifest in eq.~\eCintegral.

Now we are ready to examine eq.~\ediffii\ further.
By noting that the outward unit normal vector
$N^\mu (r)$ is related to $v^\mu (r)$ by
\eqn\eNv{N_\mu (r) =
{1 \over \rho(r,P)}~v_\alpha (r) V^\alpha_{~\mu} (P,r)~,}
and using the Taylor expansion
\eqn\eTayloru{u_\nu (r) = \sum_{m=0}^\infty
{1 \over m!}~v^{\mu_1} (r) ... v^{\mu_m} (r) \cdot \nabla_{\mu_1} ...
\nabla_{\mu_m} u_\alpha (P) \cdot V^{\alpha}_{~\mu} (P,r)~,}
we can write the second surface integral of eq.~\ediffii\ as
\eqn\esurface{\eqalign{&\int_{\rho(r,P) = \ep} \dO_\ep~N_\mu (r)
u_\nu (r) \H^\mn (r) \Phi_a (P) \cr
&\quad = \sum_{m=0}^\infty {1 \over m!} \nabla_{\mu_1} ...
\nabla_{\mu_m} u_\beta (P) \cdot
\left[ \Ct^{\alpha\beta,~~\mu_1 ... \mu_m}_{~~~~\alpha}
(\ep; h(P),g) \right]_a^{~b} \Phi_b (P) + {\rm o} (1)~,\cr}}
where we define the coefficients $\Ct$ by
\eqn\eCtilde{\eqalign{&{1 \over \ep} \int_{\rho(r,P) = \ep}
\dO_\ep~v_\alpha (r) v^{\mu_1} (r) ...
v^{\mu_m} (r) V^\alpha_{~\mu} (P,r)
V^\beta_{~\nu} (P,r) \H^\mn (r) \Phi_a (P) \cr
&\quad = \left(\Ct^{\alpha\beta, ~~\mu_1 ... \mu_m}_{~~~~\alpha} (\ep;h(P),g)
\right)_a^{~b} \Phi_b (P)
+ {\rm o} (1)~, \cr}}
in which we ignore terms that vanish in the limit $\ep \to 0$.

The conservation law \econserv\ implies that
\eqn\eintermedi{\eqalign{&\int_{\rho_1}^{\rho_2}
d\rho \int \dO_\rho ~\nabla_\mu u_\nu (r)
\H^\mn (r) \Phi_a (P) \cr
&\quad =
\left( \int \dO_{\rho_2} - \int \dO_{\rho_1} \right)
N_\mu (r) u_\nu (r) \H^\mn (r) \Phi_a (P)~.\cr}}
Hence, by differentiating this with respect to $\rho_2$
(and replacing $\rho_2$ by $\rho$),
we obtain
\eqn\eintermedii{{\partial \over \partial \rho} \int \dO_\rho ~N_\mu (r)
u_\nu (r)~\H^\mn (r) \Phi_a (P)
= \int \dO_\rho~\nabla_\mu
u_\nu (r)~\H^\mn (r) \Phi_a (P)~.}
We extract the part of eq.~\eintermedii\ that cannot
be integrated over $\rho$ up to $0$, and we obtain, from
\esurface\ and \eOPE,
\eqn\edrhoCt{{\p \over \p \rho}
\left( u \cdot \Ct (\rho; h,g) \right)_a^{~b} \Phi_b (P)
= \left(  \L_u h \cdot
\C (\rho; h,g) \right)_a^{~b} \Phi_b (P)~,}
where
\eqn\euCt{\left( u \cdot \Ct (\rho; h,g) \right)_a^{~b}
\equiv \sum_{m=0}^\infty
{1 \over m!} \nabla_{\mu_1} ... \nabla_{\mu_m} u_\beta (P)
\cdot \left(
\Ct^{\alpha\beta,~~\mu_1 ... \mu_m}_{~~~~\alpha}
(\rho; h, g) \right)_a^{~b}~.}
Eq.~\edrhoCt\ implies
that the $\rho$ dependence of the coefficients $\Ct$
is determined by the coefficients $\C$ which are
themselves determined by eq.~\esing\ in terms of
the connection $\K$.

We can now rewrite eq.~\ediffii\ using eqs.~\esurface\
and \edrhoCt\ as
\eqn\ediffiii{\eqalign{
&\vvev{\Phi_a (P)}_{h + \L_u h, g} - \vev{\Phi_a (P)}\cr
&= \left[ u \cdot \Ct (1;h,g)
- \L_u h \cdot \K (h,g)
\right]_a^{~b} \vev{\Phi_b (P)}~. \cr}}
Therefore, the consistency between the diffeomorphism
\ediffeoPhi\ and eq.~\ediffiii, which is a result of
the Ward identity \eWard, gives the second main result
of this paper:
\eqn\ecomm{\left(
u \cdot \Ct (1;h,g) \right)_a^{~b} \Phi_b (P)
 = \L_u \Phi_a (P) +
\left( \L_u h \cdot \K (h,g) \right)_a^{~b} \Phi_b (P)~.}
This can be regarded as the initial condition for the differential
equation \edrhoCt, which we will solve in the next
section.

\newsec{Further discussion of the two main results}

Eqs.~\esing\ and \ecomm\ constitute
two main results of this paper.  They express
the short-distance singularities $\C$, $\Ct$
in terms of the connection $\K$ which was introduced
as finite counterterms in the exact Ward identity
\eWard.

First we determine the $\rho$ dependence of the
coefficient $\C$, which can be obtained by solving
the RG equation \eCRG\ using \esing\ as the initial condition.
The solution is
\eqn\emainone{\delta h \cdot \C (\rho; h,g)
= {\p \over \p \rho} \left( \delta h \cdot S(\rho; h,g)
\right) ~,}
where we define
\eqn\eS{\eqalign{\delta h \cdot S(\rho; h,g) &\equiv
\Bigg[~ G(\rho; h,g) \cdot \left\{
{\delta h \over \rho^2} \cdot
\K \left( h/\rho^2, g(\ln \rho) \right) \right\}\cr
&\quad +
G(\rho; h,g) - G(\rho; h+\delta h,g) ~\Bigg]
\cdot G^{-1} (\rho;h,g)~.\cr}}
Here the matrix $G$ is defined by
\eqn\eG{\eqalign{\dt G (\rho; h,g) &\equiv
{1 \over \Delta t}
\left[ G(\e^{- \Delta t} \rho; \e^{- 2 \Delta t} h, g
+ \Delta t \beta) - G(\rho; h,g) \right] \cr
& = \gamma (h,g) G(\rho; h,g)~,
\quad G(1;h,g) = {\bf 1}~,\cr}}
and the running parameter is defined by
\eqn\erunning{{\p \over \p t}~ g^i (t) = \beta^i (g(t))~,\quad
g^i (0) = g^i~.}

Since the coefficients $\C (\rho)$ are given as
derivatives with respect to $\rho$, we can rewrite the
Ward identity \eWard\ as
\eqnn\eWardtwo
$$
\eqalignno{&\vev{\Phi_{a_1} (P_1) ... \Phi_{a_n} (P_n)}
- \vvev{\Phi_{a_1} (P_1) ... \Phi_{a_n} (P_n)}_{h + \delta h, g}\cr
& = \lim_{\epsilon \to 0} \Bigg[
\int_{\rho (r,P_k) \ge \epsilon} d^D r \sqrt{h} ~{1 \over 2} \delta
h_\mn (r) \cr
& \qquad \qquad \qquad
\times \vev{ \left( \H^\mn (r) - \vev{\H^\mn (r)} \right)
\Phi_{a_1} (P_1) ... \Phi_{a_n} (P_n)} &\eWardtwo\cr
& + \sum_{k=1}^n
\left[ \delta h (P_k) \cdot
S(\ep;h(P_k), g) \right]_{a_k}^{~b}
\vev{\Phi_{a_1} (P_1) ...  \Phi_b (P_k) ... \Phi_{a_n} (P_n)}
\quad\Bigg]~.\cr}
$$

Now, we determine the $\rho$ dependence of
the coefficients $\Ct$ by solving the differential equation
\edrhoCt\ using \ecomm\ as the initial condition.  The
solution is
\eqn\emaintwo{\left( u \cdot \Ct (\rho;h,g) \right)
\Phi = \L_u \Phi +
\left(\L_u h \cdot S(\rho; h,g) \right)
\Phi~.}
Going back to the original definition \esurface\ of the
coefficients $\Ct$, we find that eq.~\emaintwo\ gives
\eqn\eschwinger{\eqalign{&\int_{\rho(r,P) = \ep} \dO_\ep~N_\mu (r)
u_\nu (r) \H^\mn (r) \Phi_a (P)  \cr
& \quad \quad=
\L_u \Phi_a (P) + \left[ \L_u h (P) \cdot S(\ep; h(P),g) \right]_a^{~b}
\Phi_b (P) + {\rm o} (1)~.\cr}}
This surface integral is nothing but
the euclidean version of
the regularized commutator between the
energy-momentum tensor and the field
$\Phi_a$.  The first term, which gives the
change of the field $\Phi_a$ under an infinitesimal
diffeomorphism, is the canonical contribution
to the commutator, and the second term is
the Schwinger term.  Hence,
the connection $\K$ determines the
anomaly in the commutator.

To summarize, we have rewritten the main results
\esing\ and \ecomm\ by introducing the $\rho$ dependence.
Our final results are given by eqs.~\emainone\ and \emaintwo.

\newsec{Uniqueness of the energy-momentum tensor}

We have introduced the energy-momentum tensor
$\H^\mn$ as a field that generates the changes
of the correlation functions under the corresponding
change of the metric, as given by the Ward
identity~\eWard\ (or \eWardtwo).  We wonder
what properties characterize the
energy-momentum tensor uniquely.

Let us recapitulate the properties of the energy-momentum
tensor that we have obtained from consistency
of the Ward identity~\eWard.  First, it is symmetric:
\eqn\esymH{\H^\mn = \H^{\nu\mu}~,}
second, it is conserved (see \econserv):
\eqn\econsH{\nabla_\mu \H^\mn = 0~,}
third, it satisfies the canonical RG equation (see \eEMRG)
\eqn\eRGH{\dt \H^\mn = (D+2) \H^\mn~,}
and fourth, its trace is given by (see \etrace)
\eqn\etraceH{\H = \beta^i \O_i~.}
The last condition is independent of the choice
of the parameters $g^i$, since under an arbitrary
coordinate change $g^i \to g'^i$ we find
\eqn\eObetachange{\O_i \to \O_i' = {\p g^j \over \p g'^i}~\O_j~,
\quad \beta^i \to \beta'^i = {\p g'^i \over \p g^j}~\beta^j~.}
Thus, if there is any ambiguity in $\H^\mn$,
there must exist a field $\delta \H^\mn$
which satisfies
\eqn\eDH{\delta \H^\mn = \delta \H^{\nu\mu}~,
\quad \nabla_\mu \delta \H^\mn = 0~,
\quad \dt \delta \H^\mn = (D+2) \delta \H^\mn~,
\quad h_\mn \delta \H^\mn = 0~.}
We assume that such $\delta \H^\mn$ does not
exist.  Then, the four
conditions \esymH, \econsH, \eRGH, and \etraceH\ characterize
the energy-momentum tensor uniquely.

\newsec{Examples from $\phi^4$ theory}

In this section we examine the energy-momentum tensor
for the $\phi^4$ theory in four dimensions.  For simplicity
we will restrict ourselves to the flat space \eflat.
The energy-momentum tensor for this theory has been
discussed by Brown using dimensional regularization \ref\rbrown{
L.~Brown, {\it Ann. Phys.} {\bf 126}(1980)135}.
See ref.~\ref\rphifour{H.~Sonoda,
{\it Nucl. Phys.} {\bf B394}(1993)302}
for discussions of the conjugate fields in the $\phi^4$ theory.
We will do two things in this section:
first we will enumerate three conditions that specify
the energy-momentum tensor $\H^\mn$ uniquely,
and second we will determine the singularities in the product of
$\H^\mn$ and the elementary field $\phi$.

We recall that in flat space the theory is specified by three
parameters: $\lambda$, $m^2$, and $\gone$.
These satisfy the RG equations:
\eqn\eRGphifour{\eqalign{
&\dt \lambda = \beta_\lambda (\lambda) \cr
&\dt m^2 = (2 + \beta_m (\lambda)) m^2\cr
&\dt \gone = 4 \gone + {1 \over 2}~m^4 \beta_{\bf 1}
(\lambda)~.\cr}}
The parameter $\gone$ is the cosmological constant.

Let us now impose the first three conditions
\esymH, \econsH, and \eRGH.  We get
\eqnn\esym
\eqnn\econservflat
\eqnn\eEMRGflat
$$
\eqalignno{
\H^\mn &= \H^{\nu\mu}~,&\esym\cr
\p_\mu \H^\mn &= 0~,&\econservflat\cr
\dt \H^\mn &= 4 \H^\mn~.&\eEMRGflat\cr}
$$
We note that eq.~\eEMRGflat\ for the flat space
follows from eq.~\eRGH\ for the curved space
because of eq.~\eRGflat. ($m=0$, $n=2$)

We wish to show that the three conditions
\esym, \econservflat, and \eEMRGflat\ specify
$\H^\mn$ without ambiguity.  In ref.~\rbrown\ only
eqs.~\esym\ and \econservflat\ were considered.
But these two conditions leave a well-known
ambiguity in $\H^\mn$.  Namely, if $\H^\mn$
satisfies eqs.~\esym\ and \econservflat, then
\eqn\eredef{\H'^\mn \equiv \H^\mn + f(\lambda) \left(
\p^\mu \p^\nu - \delta^\mn \p_\alpha \p^\alpha \right)
\O_m~,}
where $\O_m$ is the field conjugate to $m^2$,
also satisfies the two conditions.
But the conjugate field $\O_m$, which is the same
as the renormalized $\phi^2/2$, satisfies \rbrown\ \rphifour\
\eqn\eOmRG{\dt \O_m = (2 - \beta_m (\lambda)) \O_m
- m^2 \beta_{\bf 1}~.}
Hence, if $\H^\mn$ satisfies eq.~\eEMRGflat\ then
$\H'^\mn$ defined by \eredef\ does not satisfy it for any
regular function $f(\lambda)$.  Therefore, the three conditions
\esym, \econservflat, and \eEMRGflat\ specify
the energy-momentum tensor uniquely.  Here
we did not use the trace condition \etraceH\
to prove uniqueness, but alternatively we can prove uniqueness
by showing the absence of a conserved
symmetric traceless tensor which transforms canonically under
the RG.  In appendix we will
give an explicit form of the energy-momentum tensor using
dimensional regularization.

In ref.~\rccj\ it was argued that
the energy-momentum tensor can be specified uniquely
if we demand that it be coupled to external gravity.
Imposing this condition is equivalent
to defining the energy-momentum
tensor through the Ward identity \eWard.  The energy-momentum
tensor $\H^\mn$ that satisfies the three conditions
\esym, \econservflat, and \eEMRGflat\ are obtained from
the energy-momentum tensor in curved space by taking the
limit of the flat metric.\foot{On curved space we can introduce
a new dimensionless parameter
$\eta$, whose conjugate field
is $R~\O_m$, where
$R$ is the Ricci curvature.  We must take
$\eta = 0$ before we take the flat metric limit.}

We now discuss the short-distance singularities
in the product $\H^{\mu\nu} (r) \phi (P)$.  The only relevant
matrix element of the connection is
$\left(\K^\mn (\delta, \lambda)\right)_\phi^{~\phi}$.  By covariance
we must find it in the form:
\eqn\eKphi{\left(\K^\mn (\delta, \lambda)\right)_\phi^{~\phi}
= \delta^\mn K(\lambda)~.}
The trace condition \etraceK\ implies that
\eqn\eK{K (\lambda) = {1 \over 4} \Psi_\phi^{~\phi} (\lambda)~,}
where
\eqn\ePsi{\Psi_\phi^{~\phi} (\lambda) \equiv 1 +
\gamma_\phi (\lambda) + \beta_\lambda (\lambda)
\left( c_\lambda \right)_\phi^{~\phi} (\lambda) ~.}
Here $1 + \gamma_\phi (\lambda)$ is the full scale
dimension of $\phi$, and $c_\lambda$ is the connection
in the $\lambda$ direction.
Under the redefinition of the field $\phi$
by
\eqn\eredefphi{\phi \to N (\lambda) \phi~,}
the anomalous dimension $\gamma_\phi$ and
the connection $\left( c_\lambda \right)_\phi^{~\phi}$ transform
as
\eqn\egctransform{\eqalign{
\gamma_\phi &\to \gamma_\phi + \beta_\lambda \p_\lambda
\ln N\cr
\left( c_\lambda \right)_\phi^{~\phi}
&\to \left( c_\lambda \right)_\phi^{~\phi} - \p_\lambda \ln N ~.}}
Hence, $\Psi_\phi^{~\phi}$ is invariant under the redefinition
\eredefphi.

We apply eqs.~\emainone\ and \emaintwo\
to the product
$\H^\mn (r) \phi (P)$, and obtain
\eqnn\eCphi
\eqna\eCtphi
$$
\eqalignno{&\left(\C^\mn\right)_\phi^{~\phi} (\rho; \lambda)
= {1 \over \rho}~\delta^\mn \beta_\lambda (\lambda (\ln \rho))
K' (\lambda (\ln \rho))&\eCphi\cr
&\left(\Ct^{\alpha\nu,}_{~~~~\alpha}\right)_\phi^{~\p_\mu \phi}
(\rho ; \lambda) = \delta^\mn &\eCtphi a\cr
&\left(\Ct^{\alpha\nu,~~\mu}_{~~~~\alpha}\right)_\phi^{~\phi}
(\rho; \lambda) = \delta^\mn K(\lambda(\ln \rho)) ~.&\eCtphi b\cr}
$$
If we adopt a particular scheme in which
\eqn\escheme{\left(c_\lambda\right)_\phi^{~\phi} (\lambda)
= 0~,}
then
\eqn\egPsi{K(\lambda) =
{1 \over 4}~\Psi_\phi^{~\phi} (\lambda) =
{1 \over 4}~\left( 1 + \gamma_\phi (\lambda) \right)~,}
and the OPE coefficients are completely determined by
the full scale dimension $1 + \gamma_\phi$ as in
eqs.~\eCphi\ and \eCtphi{}.  This result, to order
$\lambda^2$, was obtained a long time ago in refs.~\rcj\
and \rwilson.

\newsec{Concluding remarks}

In this paper we have studied the energy-momentum
tensor in field theory on curved space.
We have introduced the energy-momentum
tensor through an exact Ward identity \eWard\ (or \eWardtwo).
We have found that
the singular part of the OPE of the energy-momentum tensor
and an arbitrary composite field is determined
in terms of a connection
$\K$ as in eqs.~\emainone\ and \emaintwo.

Our Ward identity \eWard\ is a generalization
of the Ward identity for two dimensional
conformal field theory given in
refs.~\ref\rsonodaconf{H.~Sonoda, {\it Nucl. Phys.} {\bf B281}(1987)546; {\bf
B311}(1988/89)401} and
\ref\reo{T.~Eguchi and H.~Ooguri, {\it Nucl. Phys.} {\bf
B282}(1987)308}.  The OPE
of the energy-momentum tensor and an arbitrary
composite field is completely determined
by the conformal symmetry, and the connection
$\K$ is calculable.  In fact,
the OPE of two energy-momentum
tensors (i.e.,  the central charge) and the normalization
of three-point functions
give enough data to construct all correlation
functions \ref\rbpz{A.~Belavin, A.~Polyakov, and
A.~Zamolodchikov, {\it Nucl. Phys.} {\bf B241}(1984)333}.
This feature will not generalize to field theories
in higher space dimensions, either massless or
massive.

Another important result in this paper is
the absence of anomalies in the RG equation of the
energy-momentum tensor, eq.~\eEMRG.
This result is not new.  For example, it has
played a crucial role in the
work of Curci and Paffuti \ref\rcp{G.~Curci and G.~Paffuti,
{\it Nucl. Phys.} {\bf 286}(1987)399\semi
N.~E.~Mavromatos and J.~L.~Miramontes,
{\it Phys. Lett.} {\bf B212}(1988)33} in which
the canonical RG equation of the
energy-momentum tensor was
used to derive a particular convention
of the beta functions for
the two dimensional non-linear sigma model
in their discussion of string field equations.

Our discussion of the short-distance singularities
in sect.~6 is not complete unless we compute
the connection $\K$; our main results
\emainone, \emaintwo\ give relations between
the short-distance singularities and the counterterms
in the exact Ward identity \eWard, but
$\K$, being independent of the beta functions and
anomalous dimensions, needs to be computed
separately.  To find the connection $\K$, it helps to know
the constraints on its matrix elements.
We found one constraint on the trace of the
connection, eq.~\etraceK, in sect.~4.
There are additional algebraic constraints
as can be seen as follows.  First,
eqs.~\eCintegral\ and \eCtilde\
imply that the following conditions must be satisfied
upon contraction of indices:
\eqn\eCpartialtrace{\eqalign{
\C^{\mu\nu, \mu_1 ... \mu_m} (\rho; h,g)
&= {1 \over \rho^2}~h_{\mu_{m+1} \mu_{m+2}}
\C^{\mu\nu, \mu_1 ... \mu_{m+2}}
(\rho; h, g) + {\rm o} \left( {1 \over \rho^3} \right) \cr
\Ct^{\mu\nu,~~\mu_1 ... \mu_m}_{~~~~\mu} (\rho; h,g)
&= {1 \over \rho^2}~h_{\mu_{m+1} \mu_{m+2}}
\Ct^{\mu\nu,~~ \mu_1 ... \mu_{m+2}}_{~~~~\mu}
(\rho; h, g) + {\rm o} \left( {1 \over \rho^2} \right)~. \cr}}
Second, we note that
eq.~\emaintwo\ determines $\Ct$ with its indices partially
contracted.  There must exist uncontracted
coefficients $\Ct$ that satisfy
\eqn\eCtpartialtrace{\Ct^{\mu\nu,~~\mu_1 ... \mu_m}_{~~~~\mu} (\rho; h,g) =
h_{\mu \mu_{m+1}} \Ct^{\mu\nu,\mu_{m+1} \mu_1 ...
\mu_m} (\rho;h,g)~.}
Finally, the definitions \eCintegral\ and \eCtilde\ imply that
$\Ct$ must be related to $\C$ by
\eqn\eCtC{\Ct^{\mu\nu,\mu_1 ... \mu_m} (\rho; h,g)
= {1 \over \rho}~\C^{\mu\nu,\mu_1 ... \mu_m}
(\rho; h,g) + {\rm o} \left( {1 \over \rho^2} \right)~.}
The connection $\K$ is constrained by
the above three algebraic conditions.  So far we
have not found any simple way of rewriting these
constraints as constraints directly on $\K$.

In part II and III of the present paper we plan to discuss the
torsion $\tau$ and curvature $\Omega$
of the connection $\K$, respectively:
\eqnn\etorsion
\eqnn\ecurv
$$
\eqalignno{
&\tau (\delta h_1, \delta h_2; h,g) \equiv
\left( \delta h_1 \cdot \K \right) (\delta h_2)_{\alpha
\beta} \H^{\alpha\beta} -
\left( \delta h_2 \cdot \K \right) (\delta h_1)_{\alpha
\beta} \H^{\alpha\beta} ~,&\etorsion\cr
& \Omega (\delta h_1, \delta h_2; h,g)
\equiv \delta h_2 \cdot \left(\K (h + \delta h_1,g)
- \K (h,g) \right) \cr
&~~ - \delta h_1 \cdot \left(\K (h + \delta h_2,g)
- \K (h,g) \right)  + \left[ \delta h_1 \cdot
\K (h,g) , \delta h_2 \cdot \K (h,g) \right]~.
&\ecurv \cr}
$$

\bigbreak\bigskip\bigskip\centerline{
{\bf Acknowledgment}}\nobreak
I thank Barton Zwiebach for his interest in and suggestions
for this work.

\appendix{A}{Construction of the energy-momentum tensor
in $\phi^4$ theory using dimensional regularization}

In this appendix we construct the energy-momentum
tensor for $\phi^4$ theory using dimensional regularization.
For the most part we follow ref.~\rbrown.

The theory is defined perturbatively in $D = 4 - \ep$
dimensional euclidean space by the lagrangian
\eqn\elag{\L = {1 \over 2} \p^\mu \phi_0 \p_\mu \phi_0
+ m_0^2 {\phi_0^2 \over 2}
+ \lambda_0 {\phi_0^4 \over 4!}
+ g_0~,}
where the bare parameters are given in terms of the renormalized
parameters $\lambda$, $m^2$, and $\gone$ as
\eqn\epara{
\lambda_0 = Z_\lambda (\ep;\lambda) \lambda~,\quad
m_0^2 = Z_m (\ep; \lambda) m^2~,\quad
g_0 = \gone + z_0 (\ep; \lambda) {m^4 \over 2}~.}
We adopt the MS scheme:  $Z_\lambda - 1$, $Z_m - 1$,
and $z_0(\ep;\lambda)$ all contain only the pole
part with respect to $\ep$.
We fix the usual arbitrary scale $\mu^2$ at $1$ for simplicity.
The renormalization constants are related to the
beta functions (see eqs.~\eRGphifour) as
\eqn\ebeta{\eqalign{
\ep \lambda + \beta_\lambda (\lambda) &=
{\ep \lambda Z_\lambda \over \p_\lambda
(\lambda Z_\lambda)} \cr
\beta_m (\lambda) &= - (\ep \lambda + \beta_\lambda)
\p_\lambda \ln Z_m\cr
\beta_{\bf 1} (\lambda) &= - (\ep \lambda + \beta_\lambda)
\p_\lambda z_0 - (\ep + 2 \beta_m) z_0~.\cr}}
$\beta_\lambda$, $\beta_m$, and $\beta_{\bf 1}$
are of order $\lambda^2$, $\lambda$, and $1$, respectively.

The renormalized composite fields are given as follows:
\eqnn\ephifour
\eqnn\ephitwo
$$
\eqalignno{\left[ {\phi^4 \over 4!} \right]
&\equiv \p_\lambda \left( \lambda Z_\lambda \right)
\cdot {\phi_0^4 \over 4!} + m^2 \p_\lambda Z_m \cdot
{\phi_0^2 \over 2} + {m^4 \over 2} \p_\lambda z_0
+ z_R (\ep;\lambda) \p^2 {\phi_0^2 \over 2}~,&\ephifour\cr
\left[{\phi^2 \over 2}\right] &\equiv Z_m ~{\phi_0^2 \over 2}
+ z_0 m^2~.&\ephitwo\cr}
$$
Here $z_R (\ep;\lambda)$ contains only the pole part,
and it is of order $\lambda^2$ due to the Feynman diagram
shown in Fig. 2.

\bigskip
\centerline{\epsfxsize=0.7\hsize \epsfbox{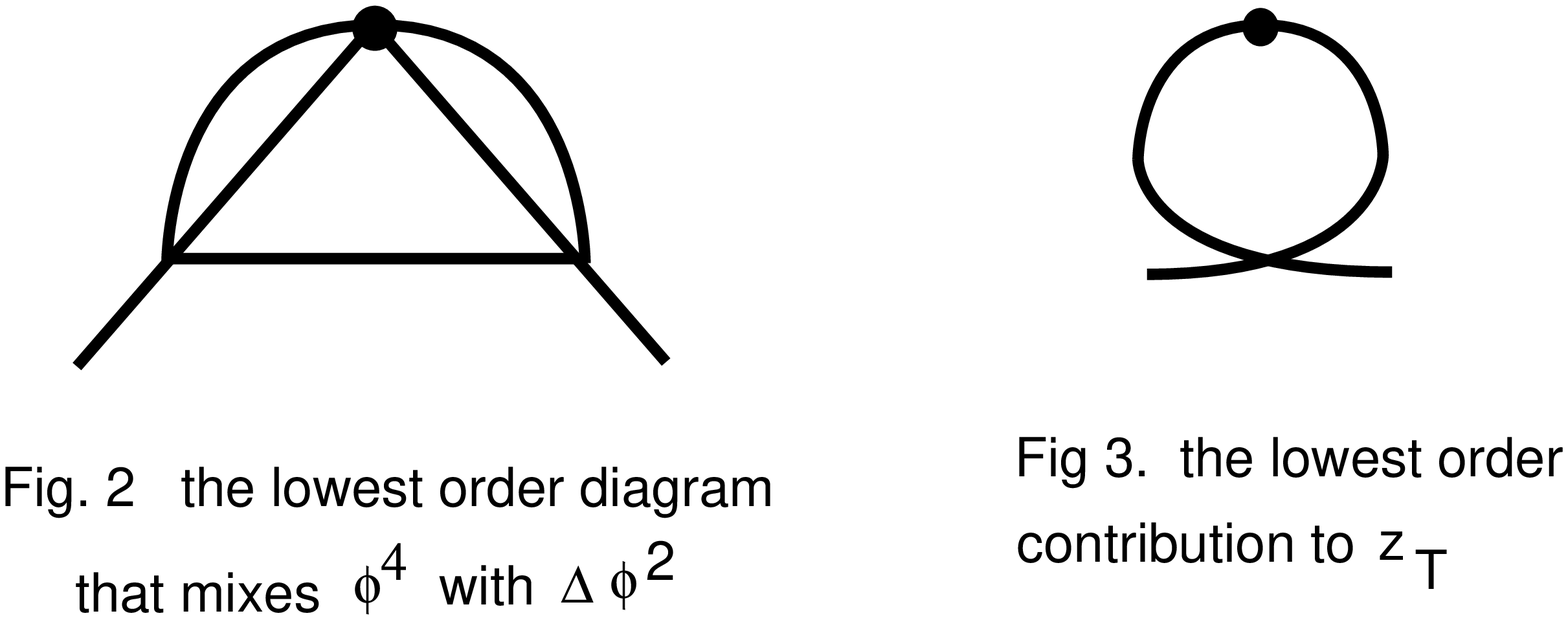}}

We can find the RG equations satisfied by these
composite fields in the usual way \ref\rthooft{G.~'t~Hooft,
{\it Nucl. Phys.} {\bf B62}(1973)444} as
\eqnn\eRGphifour
\eqnn\eRGphitwo
$$
\eqalignno{\dt \left[ {\phi^4 \over 4!}\right]
&= (4 - \beta'_\lambda) \left[ {\phi^4 \over 4!}\right]
- \beta_m' m^2 \left[{\phi^2 \over 2}\right] - \beta_{\bf 1}'
{m^4 \over 2} + u_R (\lambda) \p^2 \left[ {\phi^2 \over 2}\right]
&\eRGphifour\cr
\dt \left[{\phi^2 \over 2}\right] &=
(2-\beta_m) \left[{\phi^2 \over 2}\right] - \beta_{\bf 1} m^2~,
&\eRGphitwo\cr}
$$
where $u_R$ is defined by
\eqn\euR{u_R (\lambda) \equiv {\p_\lambda \left(
(\ep \lambda + \beta_\lambda) z_R \right) \over
Z_m}~,}
and it is of order $\lambda^2$.

The general formula \eRGconj\ implies that
the fields conjugate to $\lambda$, $m^2$ must satisfy
the RG equations
\eqnn\eRGOlhere
\eqnn\eRGOmhere
$$
\eqalignno{\dt \O_\lambda &= (4 - \beta_\lambda') \O_\lambda
- m^2 \beta_m' \O_m - {m^4 \over 2}~\beta_{\bf 1}'~,&\eRGOlhere\cr
\dt \O_m &= (2 - \beta_m) \O_m - m^2 \beta_{\bf 1}~.
&\eRGOmhere \cr}
$$
Eqs.~\eRGphitwo\ and \eRGOmhere\ imply
\eqn\eOmphitwo{\O_m = \left[ {\phi^2 \over 2}\right]~.}
On the other hand, eqs.~\eRGphifour\ and
\eRGOlhere\ imply that the renormalized $\phi^4/4!$
differs from $\O_\lambda$ by a total derivative:
\eqn\eOlphifour{\O_\lambda = \left[ {\phi^4 \over 4!}\right]
+ f(\lambda) \p^2 \O_m~,}
where $f(\lambda)$ satisfies
\eqn\ef{
\p_\lambda ( \beta_\lambda f ) - \beta_m f + u_R = 0~.}
Eq.~\ef\ has a unique solution which is regular
at $\lambda = 0$.  The general formula \etraceH\
gives the trace of the energy-momentum
tensor as
\eqn\etracephifour{\H =
\beta_\lambda \O_\lambda + (2 + \beta_m) m^2 \O_m +
4 \gone + {m^4 \over 2}~\beta_{\bf 1}~.}

In ref.~\rbrown\ it was shown that we can construct two
independent traceless symmetric tensors of dimension four:
\eqnn\eT
\eqnn\et
$$
\eqalignno{T^\mn &\equiv \phi_0 \left(\p^\mu \p^\nu - {1
\over D} \delta^\mn \p^2 \right) \phi_0+
z_T (\ep;\lambda)\left(\p^\mu \p^\nu - {1
\over D} \delta^\mn \p^2 \right) {\phi_0^2 \over 2}&\eT\cr
t^\mn &\equiv \left(\p^\mu \p^\nu - {1
\over 4} \delta^\mn \p^2 \right) \O_m~,&\et\cr}
$$
where $z_T$ includes only the pole part, and it is of order
$\lambda$.  See Fig. 3.
The traceless tensors $T^\mn$, $t^\mn$ satisfy the
RG equations:
\eqnn\eTRG
\eqnn\etRG
$$
\eqalignno{\dt T^\mn &= 4 T^\mn + \eta (\lambda) t^\mn
&\eTRG\cr
\dt t^\mn &= (4 - \beta_m) t^\mn ~,&\etRG\cr}
$$
where
\eqn\eeta{\eta (\lambda) \equiv {\left( \ep \lambda + \beta_\lambda
\right) \p_\lambda z_T \over Z_m}}
is of order $\lambda$.

The traceless part of the energy-momentum tensor
$\H^\mn - {1 \over 4} \delta^\mn \H$ must be a linear
combination of $T^\mn$ and $t^\mn$:
\eqn\eHTt{\H^\mn - {1 \over 4} \delta^\mn \H
= a(\lambda) T^\mn + b(\lambda) t^\mn~.}
We can determine
the coefficients $a$, $b$ by demanding the conservation law
\eqn\econservphifour{\eqalign{&\p_\mu \left(
a(\lambda) T^\mn + b(\lambda) t^\mn \right) =
- {1 \over 4} \p^\nu \H \cr
& \qquad = - {1 \over 4} \left(
\beta_\lambda \left[ {\phi^4 \over 4!} \right]
+ (2+\beta_m) m^2 \O_m + \beta_\lambda f \p^2 \O_m \right)~.\cr}}
We find, from \eT, that
\eqn\edivT{\p_\mu T^\mn =
- {1 \over 4} \p^\nu \left( \beta_\lambda \left[
{\phi^4 \over 4!} \right] + (2 + \beta_m) m^2 \O_m \right)
+ {1 \over 4}~\chi~\p^\nu \p^2 \O_m~,}
where
\eqn\echi{\chi (\lambda) \equiv
{1 \over Z_m}
\left( (\ep \lambda + \beta_\lambda) z_R
+ 2 + {\ep \over 2} \left( Z_m - 1 \right) + z_T (3 - \ep)\right)~.}
We also find trivially
\eqn\edivt{\p_\mu t^\mn = {3 \over 4} \p^\nu \p^2 \O_m~.}
Hence, the condition \econservphifour\ determines
\eqn\eab{a = 1~,\quad
b = - {1 \over 3}~\left( \beta_\lambda f + \chi \right)~.}

We can actually show that $\chi$ and $u_R$ are determined
by $\eta$ as follows.  By differentiating eq.~\echi\
with respect to $\lambda$, we obtain
\eqn\echirelation{
(\ep \lambda + \beta_\lambda) \p_\lambda \chi - \beta_m \chi
= (\ep \lambda + \beta_\lambda) u_R - {\ep \over 2} \beta_m
+ (3-\ep) \eta~,}
where we used eqs.~\ebeta, \euR, and \eeta.
Eq.~\echirelation\ has two kinds of
terms: those zeroth order in $\ep$ and those
first order in $\ep$.
By taking the zeroth order terms, we obtain
\eqn\echizero{(\beta_\lambda \p_\lambda - \beta_m ) \chi
= \beta_\lambda u_R + 3 \eta~,}
and by taking the first order terms, we obtain
\eqn\echione{\lambda \p_\lambda \chi = \lambda u_R
- {1 \over 2}~\beta_m - \eta~.}
We find, from these two equations,
\eqn\echieta{\chi = {3 \over 2}
- \left(3 + {\beta_\lambda \over \lambda}
\right) \sigma~,}
and
\eqn\euReta{u_R = - 3~\p_\lambda \sigma
- \p_\lambda
\left({\beta_\lambda \over \lambda}~\sigma \right)
+ {\beta_m \over \lambda}~\sigma~,}
where we define
\eqn\esigma{\sigma \equiv
{1 \over 2} +{\eta \over \beta_m}~.}
Eqs.~\echieta\ and \euReta\ determine
$\chi$ and $u_R$ in terms of $\eta$ and $\beta_m$.
Since eq.~\echi\ implies
\eqn\echiatzero{\chi (0) = 2~,}
we obtain, from eq.~\echieta,
\eqn\esigmaatzero{
\sigma (0) = - {1 \over 6}~.}
Similarly, we can obtain $\sigma'(0)$, $\sigma''(0)$
from eq.~\euReta\
by recalling that $u_R$ is of order $\lambda^2$.

Finally, we verify that the
traceless tensor \eHTt\ has no anomaly
under the RG:
\eqn\eRGphifour{\dt \left(
T^\mn + b(\lambda) t^\mn \right)
= 4 \left(
T^\mn + b(\lambda) t^\mn \right)~.}
This would imply
\eqn\eanother{\left( \beta_\lambda \p_\lambda - \beta_m \right)
b + \eta = 0~.}
Using \eab, this condition is equivalent to
\eqn\estillanother{\left( \beta_\lambda \p_\lambda -
\beta_m \right) ( \beta_\lambda f + \chi ) - 3 \eta = 0~,}
which is indeed satisfied thanks to eqs.~\ef\ and \echizero.

To summarize, we have found the traceless part
of the energy-momentum tensor as eq.~\eHTt,
where $a$, $b$ are given by eqs.~\eab.

\listrefs
\parindent=20pt
\bye